\documentclass[a4paper,10pt]{article}
\usepackage{amsmath,amssymb,amstext,amsfonts}
\usepackage{amsthm, mathrsfs}
\usepackage{graphicx,subfigure}
\usepackage{url}
\usepackage{authblk}
\usepackage{natbib}
\usepackage{dsfont} 

\newcommand{\naturalnumbers}{\mathbb{N}}

\newcommand{\expectation}{\mathbb{E}}
\newcommand{\variance}{\textrm{Var}}
\newcommand{\covariance}{\textrm{Cov}}
\newcommand{\stddev}{\textrm{SD}}

\newcommand{\Vco}{\textrm{Vco}}

\newcommand{\nummbrs}{N}

\newcommand{\benefit}{B}

\newcommand{\execmult}{k}
\newcommand{\execpropn}{\alpha}
\newcommand{\execfactor}{f}

\newcommand{\liability}{X}
\newcommand{\liabilityoneunit}{Y}
\newcommand{\liabilitytotal}{L}

\newcommand{\lifetable}{PMA92C10 }
\newcommand{\lifetablenospace}{PMA92C10}
\newcommand{\memberage}{40}
\newcommand{\agerating}{r}

\newcommand{\riskweight}{\pi}
\newcommand{\riskweightexec}{\riskweight^{\textrm{exec}}}
\newcommand{\riskweightnormal}{\riskweight^{\textrm{norm}}}
\newcommand{\riskalloc}{p}
\newcommand{\riskallocexec}{\riskalloc^{\textrm{exec}}}

\newcommand{\riskallocbenefit}{\tilde{p}}
\newcommand{\riskallocbenefitexec}{\riskallocbenefit^{\textrm{exec}}}

\theoremstyle{plain}

\theoremstyle{definition}

\theoremstyle{remark}

\numberwithin{equation}{section}

\title{Quantifying mortality risk in small defined-benefit pension schemes}

\author{Catherine Donnelly\thanks{Department of Actuarial Mathematics and Statistics and the Maxwell Institute for Mathematical Sciences, Heriot-Watt University, Edinburgh EH14 4AS, United Kingdom. Email: {\tt C.Donnelly@hw.ac.uk}}}

\affil{}

\begin{document}

\maketitle

\begin{abstract}
A risk of small defined-benefit pension schemes is that there are too few members to eliminate idiosyncratic mortality risk, that is there are too few members to effectively pool mortality risk.  This means that when there are few members in the scheme, there is an increased risk of the liability value deviating significantly from the expected liability value, as compared to a large scheme.

We quantify this risk through examining the coefficient of variation of a scheme's liability value relative to its expected value.  We examine how the coefficient of variation varies with the number of members and find that, even with a few hundred members in the scheme, idiosyncratic mortality risk may still be significant. 

Next we quantify the amount of the mortality risk concentrated in the executive section of the scheme, where the executives receive a benefit that is higher than the non-executive benefit.  We use the Euler capital allocation principle to allocate the total standard deviation of the liability value between the executive and non-executive sections.   The results suggest that the mortality risk of the scheme should be monitored and managed within the sections of a scheme, and not only on a scheme-wide basis.
\end{abstract}


\section{Introduction} \label{intro}

A risk of small defined-benefit pension schemes is that there are too few members to effectively pool mortality risk.  Here mortality risk refers to the risks associated with members' future lifetimes.  To define what we mean by ``pooling mortality risk'', we need to look at the components of mortality risk.

Mortality risk can be decomposed into two parts: systematic risk and idiosyncratic risk.  Systematic risk arises from our uncertainty as to the correct distribution of deaths.  For example, an actuary may decide that mortality will improve in the future but is uncertain as to the level of mortality improvement; this particular type of systematic risk is often called longevity risk (see \cite{crawfordetal08.article} for a recent literature review on the topic of longevity risk).   Idiosyncratic risk comes about through random fluctuations: we may know the correct distribution of deaths for the scheme but we don't know exactly when each member will die.  The decline of this latter risk as the number of members increases is what we mean by ``pooling mortality risk''.

The quantification of mortality risk is the main focus of the paper.  We quantify both the amount of systematic and idiosyncratic risk in a small defined-benefit pension scheme.  Furthermore, we attempt to allocate the mortality risk between the scheme members in order to identify groups of members in which mortality risk is concentrated.  This kind of analysis may help the scheme manage the mortality risk posed by some groups of members, and perhaps aid in making investment decisions such as purchasing annuities or buying-out part of the scheme's liability with an insurance company.

The quantification of the risk of small pension schemes is touched upon in \cite[Section 11.5]{richardsjones04.unpublished}, where the amount of assets required to meet at a 99\% confidence level the present value of the liabilities is calculated for a few pension schemes.  Using an extreme quantile-based risk measure, for example value-at-risk or expected shortfall, is reasonable if there is a contractual obligation to provide benefits since the benefits must be paid whichever future scenario arises.  However, we are assuming that the pension scheme's benefits are promised rather than guaranteed.  We choose the standard deviation as a risk measure as it captures ``everyday'' scenarios, rather than one which captures highly unlikely, highly adverse scenarios under which the benefits will probably not be paid in full anyway.  From a practical viewpoint, the standard deviation is easy to understand and to calculate.  As our focus is on small defined-benefit pension schemes, these are appealing properties since small schemes generally cannot afford to pay for detailed risk management services and advice (we do not discuss the ethics of such a situation).

More precisely, we quantify the mortality risk in a defined-benefit scheme by calculating the standard deviation of the present value of the scheme's liabilities as a percentage of the expected value.  This is called the \emph{coefficient of variation} and often abbreviated $\Vco$:
\begin{displaymath}
\textrm{Coefficient of variation ($\Vco$)} := \frac{\textrm{standard deviation of liabilities}}{\textrm{expected value of liabilities}}.
\end{displaymath}
The reason for considering the coefficient of variation rather than the standard deviation is due to magnitude: a standard deviation of \pounds $100 \, 000$ in the liability value is potentially a problem for a scheme with an expected liability value of \pounds $1 \, 000 \, 000$ ($\Vco=10\%$) but much less so for a scheme with an expected liability value of \pounds $10 \, 000 \, 000$ ($\Vco=1\%$).  In Section \ref{SECdiversify}, we examine how the coefficient of variation changes as we increase the number of members in the scheme for two types of mortality models: deterministic and stochastic.  When a deterministic mortality model is used, all of the mortality risk is idiosyncratic risk since using a deterministic mortality model implies that the distribution of deaths is known.  Stochastic models allow us to see the additional increase in mortality risk due to the systematic risk caused by uncertainty about the distribution of deaths.  We find that the coefficient of variation under both types of mortality model is not negligible for schemes of up to 500 members.

A related work is \cite{milevskyetal06.article}, which considers a group of longevity insurance policies and examines how the average standard deviation per policy varies as the number of policies increases.  In \cite{milevskyetal06.article}, both a deterministic and stochastic exponential model are used in the analysis and the numerical results obtained are in a similar vein to those found in this paper. 

The numerical analysis we do in the paper is performed using a simple defined-benefit pension scheme, detailed in Section \ref{SECmodelscheme}, in which all members are the same age and all receive the same constant annuity benefit upon survival to age 65.  (We can also regard this as a group of deferred annuity contracts.)  However, assuming that all members receive the same benefit amount is a simplification. 

For this reason, we examine in Section \ref{SECexecs} how an executive section in the scheme contributes to the mortality risk of the scheme.  The executives receive a multiple of the benefit paid to the non-executive members.  A significant increase in the coefficient of variation of the scheme is observed as the executives receive larger and larger benefits, although this effect diminishes as the total membership increases.  The extra funding required to meet varied benefits at a 99\% confidence level for a portfolio of 50 lives is touched upon in \cite[Section 11.5]{richardsjones04.unpublished}.  They look at three different possible benefit structures, but their analysis is not extensive or done in a systematic way as we do here.

In \cite[Section 8]{richardscurrie11.article}, longevity risk is considered for a sample defined-benefit pension scheme of over 2\,000 members.  The authors examine the distribution of the liability value around the median liability value and use this to illustrate the effect of idiosyncratic risk on the liability valuation.  They also mention that there is a concentration risk, by which they mean the risk that a large proportion of the liability is concentrated in a relatively small number of members.   However, they do not explicitly attempt to measure this risk.

Therefore, having observed the increase in the coefficient of variation caused by a small executive section, we consider in Section \ref{SECcapitalallocation} how to split the standard deviation of the liabilities between the executive and non-executive section in such a way that reflects their risks posed to the scheme.  This is a \emph{(risk) capital allocation} problem, where the risk capital is defined to be the standard deviation of the liabilities.  We use the \emph{Euler capital allocation principle} to solve this, which uses the covariance function to measure the dependencies between the members.  In the numerical examples, it is seen that the proportion of the total standard deviation allocated to the executives is much higher than the proportion of the executives who are members, and higher than using a benefit-weighted capital allocation.  Finally, we conclude in Section \ref{SECsummary}.

\section{A model of a simple defined-benefit scheme} \label{SECmodelscheme}
We consider a closed defined-benefit pension scheme which pays only a single-life deferred annuity benefit.  There are $\nummbrs$ members in the scheme and all members are the same age $x \leq 65$.  Upon survival to age 65, member $n$ receives a constant amount $\benefit_{n}$ per annum starting from age 65, with all payments made continuously.

We ignore the assets of the scheme and focus solely on the liability valuation.  The scheme liabilities are discounted using a constant interest rate and there are no increases to the benefits either before or after retirement.

Let the random variable $\liabilityoneunit_{n}$ denote the present value of 1 unit per annum payable to member $n$ upon survival to age 65.  As all members are assumed to be the same age, $Y_{1}, Y_{2}, \ldots$ are identically distributed random variables.  We assume that the expectation and variance of $\liabilityoneunit_{n}$ exist and are finite.  Using the random variable $\liability_{n}$ to denote the present value of the liability due to member $n$, the following relation holds:
\begin{displaymath}
\liability_{n} = \benefit_{n} \liabilityoneunit_{n}, \quad \textrm{for each $n=1,\ldots,\nummbrs$}.
\end{displaymath}
In standard actuarial notation, the expected value of $\liability_{n}$ is
\begin{displaymath}
\expectation ( \liability_{n} ) = \benefit_{n} \expectation ( \liabilityoneunit_{n} ) = \benefit_{n} v^{65-x} \, {}_{65-x} p_{x} \, \bar{a}_{65},
\end{displaymath}
in which $v$ is the annual discount factor, ${}_{65-x} p_{x}$ is the probability of survival from age $x$ to age $65$ and $\bar{a}_{65}$ is the expected present value of an annuity of 1 unit p.a. payable continuously from age 65 until death.
Let the random variable
\begin{displaymath}
\liabilitytotal_{\nummbrs} = \sum_{n=1}^{\nummbrs} \liability_{n}
\end{displaymath}
denote the present value of the total liability value.  In many actuarial valuations, the liability value of the scheme is calculated by taking the expected value of the present value random variable $\liabilitytotal_{\nummbrs}$, which for the scheme described is given by
\begin{displaymath}
\expectation ( \liabilitytotal_{\nummbrs} ) = \sum_{n=1}^{\nummbrs} \expectation ( \liability_{n} ) = \sum_{n=1}^{\nummbrs} \benefit_{n} \expectation ( \liabilityoneunit_{n} ) = \sum_{n=1}^{\nummbrs} \benefit_{n} v^{65-x} \, {}_{65-x} p_{x} \, \bar{a}_{65}.
\end{displaymath}
The standard deviation of $\liabilitytotal_{\nummbrs}$ is not calculated, since it is assumed that $\liabilitytotal_{\nummbrs} \approx \expectation ( \liabilitytotal_{\nummbrs} )$, a.s., for large enough $\nummbrs$.  The justification for this is first to assume that lives are independent.  Under the independence assumption,  
\begin{equation} \label{EQNvarianceliability}
\variance \left( \liabilitytotal_{\nummbrs} \right) = \sum_{n=1}^{\nummbrs} \variance \left( \benefit_{n} \liabilityoneunit_{n} \right) = \sum_{n=1}^{\nummbrs} \benefit_{n}^{2} \, \variance \left( \liabilityoneunit_{1} \right),
\end{equation}
and hence
\begin{displaymath}
\stddev \left( \liabilitytotal_{\nummbrs} \right) = \left( \sum_{n=1}^{\nummbrs} \benefit_{n}^{2} \right)^{1/2} \, \stddev \left( \liabilityoneunit_{1} \right).
\end{displaymath}
While the standard deviation increases with the number of members, this is not typically a concern.  Expressed as a percentage of the expected value, the standard deviation of the total liability value decreases as the number of members increases.  The statistic to consider to see this is the coefficient of variation
\begin{displaymath}
\Vco ( \liabilitytotal_{\nummbrs} ) = \frac{\stddev ( \liabilitytotal_{\nummbrs} )}{\expectation ( \liabilitytotal_{\nummbrs} )}.
\end{displaymath}  
Assuming that lives are independent, then
\begin{equation} \label{EQNvcoliability}
\Vco ( \liabilitytotal_{\nummbrs} ) = \frac{ \left( \sum_{n=1}^{\nummbrs} \benefit_{n}^{2} \right)^{1/2} \stddev \left( \liabilityoneunit_{1} \right)}{\sum_{n=1}^{\nummbrs} \benefit_{n} \expectation ( \liabilityoneunit_{1} ) }.
\end{equation}
Hence
\begin{displaymath}
\Vco ( \liabilitytotal_{\nummbrs} ) \leq \frac{1}{\nummbrs^{1/2}} \frac{\stddev \left( \liabilityoneunit_{1} \right)}{ \expectation ( \liabilityoneunit_{1} ) },
\end{displaymath}
and we see that the coefficient of variation converges to zero as $\nummbrs$ tends to infinity.  This means that, for large enough $\nummbrs$, the standard deviation of the total liability value is only a tiny fraction of the expected value and this justifies assuming $\liabilitytotal_{\nummbrs} \approx \expectation ( \liabilitytotal_{\nummbrs} )$, a.s.

But how large enough does $\nummbrs$ have to be to confidently say that $\Vco ( \liabilitytotal_{\nummbrs} ) \approx 0$?  If the number of members $\nummbrs$ is small, then this approximation may not be reasonable.  This is the question that motivates the paper.  

\section{Idiosyncratic and systematic risk} \label{SECdiversify}

In this section, we calculate for our model scheme how $\Vco ( \liabilitytotal_{\nummbrs} )$ varies with $\nummbrs$ under a deterministic and stochastic mortality models.  We assume that all scheme members are age $\memberage$ and set $\benefit_{n}=1$, for all $n$, so that all members receive the same amount of benefit in retirement.  The continuously-compounded interest rate is 4\% per annum.

For the deterministic model, the coefficient of variation converges to zero as the number of members in the scheme is increased.  Hence we can think of the mortality risk which exists in the scheme due it having too few members for this convergence to occur as idiosyncratic risk.  For the stochastic models, some of the mortality risk cannot be removed by increasing the number of members.  This risk which cannot be removed is called systematic risk.

\subsection{Deterministic Mortality Model}
We begin by assuming that we believe the correct distribution of deaths for the scheme to be modelled by the life table \lifetablenospace, which is based on United Kingdom pensioner male annuitant mortality (see \cite{pmatables.book}), with no age rating and a uniform distribution of deaths between integer ages, and we refer to this as the Deterministic Mortality Model.  The particular choice of the life table is somewhat arbitrary: we picked one which broadly captures human mortality rates experienced at different ages.

\begin{figure}[p]
\centering
{ \includegraphics[scale=0.7]{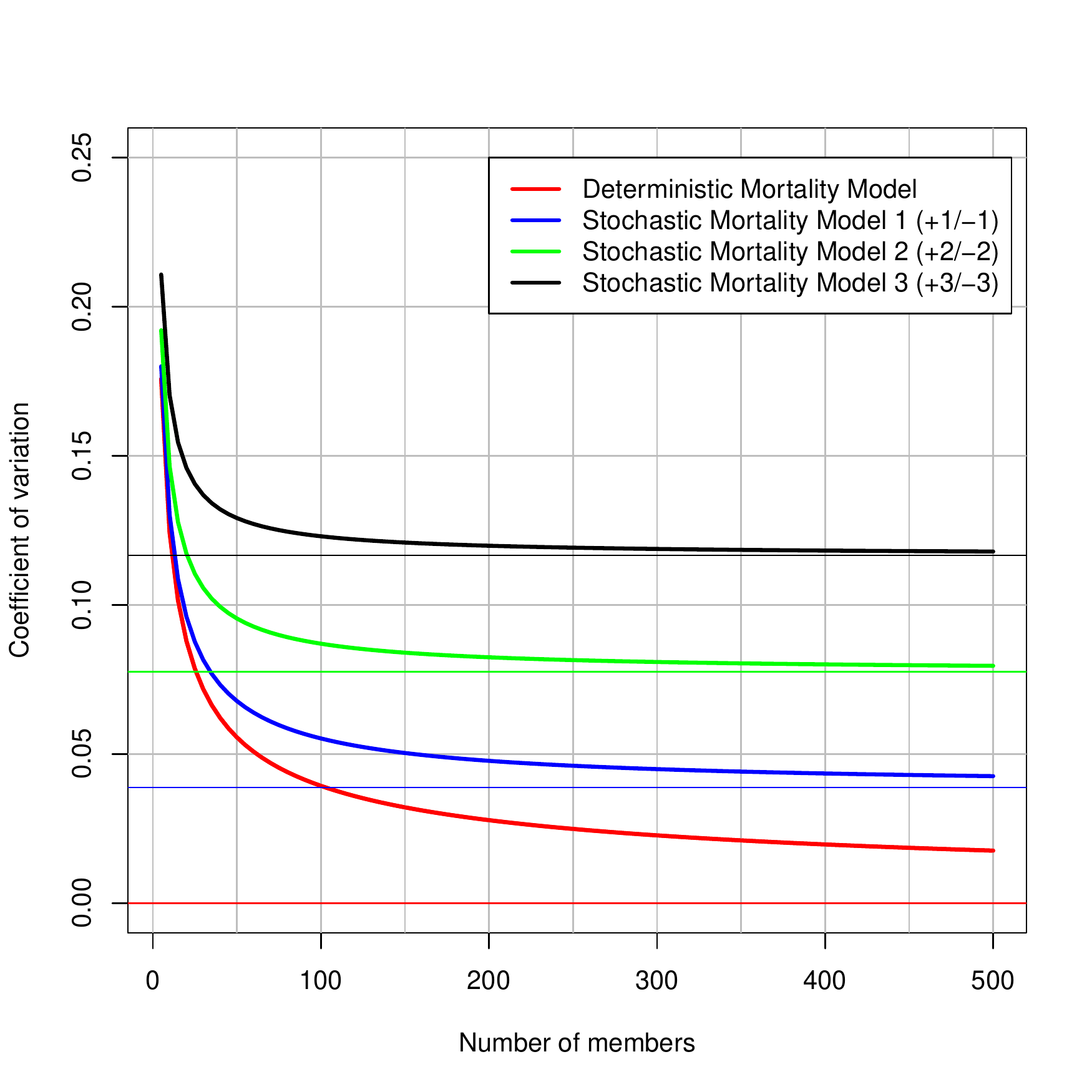} } 
\caption{Coefficient of variation for the model pension scheme using \lifetable as the basic life table under various assumptions about the age rating.  The horizontal lines show the minimum values of the coefficient of variation that can be obtained under each mortality model.}
\label{FIGVcobasic}
\end{figure}

The red curve in Figure \ref{FIGVcobasic} shows the change in the coefficient of variation as the number of members increases.
For example, when $\nummbrs=100$, $\Vco ( \liabilitytotal_{100} )=0.039$, i.e. when there are 100 members in the scheme, the standard deviation of the present value of the liabilities is 3.9\% of the expected value.  Even for a scheme with 500 members the standard deviation is still not negligible, at 1.8\% of the expected value.  However, as the number of members $\nummbrs$ becomes larger, the coefficient of variation converges to zero: this convergence limit is shown by the horizontal red line in Figure \ref{FIGVcobasic}.   The results suggest that when there are few members in the scheme, the approximation $\liabilitytotal_{\nummbrs} \approx \expectation ( \liabilitytotal_{\nummbrs} )$, a.s., which is often used by pensions actuaries, may not be appropriate.

The mortality risk under the Deterministic Mortality Model is entirely diversifiable; as the number of members increases, the mortality risk converges to zero.  The red curve in Figure \ref{FIGVcobasic} illustrates the impact of having too few members in the scheme to eliminate this idiosyncratic risk.  

\subsection{Stochastic Mortality Models} \label{SUBSECsmm}

It is common for pension actuaries to use a standard mortality table but to age rate the scheme members to reflect the members' mortality differing from the standard mortality table.   For example, age rating members down by one year (-1) means that we treat them as if they are one year younger for the purposes of reading the life table.  Correspondingly, we would also age rate down by one year the retirement age.  The overall effect is that the members must wait the same time until retirement, but they live almost one year more in retirement.  Effectively, we have extended the members' lifetimes by almost one year.

Suppose there is uncertainty about which age rating to use: perhaps we can think of solid reasons why members' life expectancies may increase (which means age rating down) but we can also think of good reasons why their life expectancies may decrease (which means age rating up).  This may be particularly relevant for small schemes because they have too few members to separate past trends in scheme mortality (due to systematic risk) from random mortality fluctuations (due to idiosyncratic risk) to reliably estimate the past distribution of deaths, let alone make predictions about the future distribution.  Thus there may be more than one age rating that a pensions actuary may consider as suitable to model the scheme mortality.  Assigning a probability to each of the possible age ratings results in a stochastic mortality model.

We use the above method to introduce a series of simple stochastic mortality models.  Fix a positive real number $\agerating$ and set
\begin{equation} \label{EQNagerating}
\textrm{age rating} =
\left\{ \begin{array}{ll}
\agerating & \textrm{with probability $0.5$}, \\
-\agerating & \textrm{with probability $0.5$}.
\end{array} \right.
\end{equation}
We call this \emph{Stochastic Mortality Model $\agerating$} and note that the difference in life expectancies between the two possible future scenarios is almost $2 \agerating$ years.  The key point to note about the Stochastic Mortality Models is that they imply that future lifetimes are no longer independent random variables.  Consequently, the  random variables $\liabilityoneunit_{1}, \liabilityoneunit_{2}, \ldots, \liabilityoneunit_{\nummbrs}$ either all have the distribution implied by \lifetable $+\agerating$ or they all have the distribution implied by \lifetable $-\agerating$, and hence they are not independent.  

Recall that $\benefit_{n}=1$ in this section.  The expected value of $\liabilitytotal_{\nummbrs}$ can still be calculated
\begin{displaymath}
\expectation_{\agerating} ( \liabilitytotal_{\nummbrs} ) = \sum_{n=1}^{\nummbrs} \benefit_{n} \expectation_{\agerating} ( \liabilityoneunit_{n} ) = \nummbrs \expectation_{\agerating} ( \liabilityoneunit_{1} ),
\end{displaymath}
where we use $\expectation_{\agerating}$ to denote the expectation under Stochastic Mortality Model $\agerating$.  The variance of $\liabilitytotal_{\nummbrs}$ is more complicated: while the random variables $\liabilityoneunit_{1}, \liabilityoneunit_{2}, \ldots$ are still identically distributed (all members are the same age), they are no longer independent.  We have
\begin{displaymath}
\variance_{\agerating} \left( \liabilitytotal_{\nummbrs} \right) = \nummbrs \variance_{\agerating} \left( \liabilityoneunit_{1} \right) + \nummbrs ( \nummbrs - 1) \covariance_{\agerating} ( \liabilityoneunit_{1}, \liabilityoneunit_{2} )
\end{displaymath}
and hence
\begin{equation} \label{EQNvcoallmbr}
\Vco_{\agerating} ( \liabilitytotal_{\nummbrs} ) = \frac{1}{\expectation_{\agerating} ( \liabilityoneunit_{1} )} \left( \frac{1}{\nummbrs} \left( \variance_{\agerating} \left( \liabilityoneunit_{1} \right) -  \covariance_{\agerating} ( \liabilityoneunit_{1}, \liabilityoneunit_{2} ) \right) + \covariance_{\agerating} ( \liabilityoneunit_{1}, \liabilityoneunit_{2} ) \right)^{1/2}.
\end{equation}
From the latter expression, we see that under Stochastic Mortality Model $\agerating$, the coefficient of variation never reaches zero but 
\begin{displaymath}
\Vco_{\agerating} ( \liabilitytotal_{\nummbrs} ) \rightarrow \frac{\left( \covariance_{\agerating} ( \liabilityoneunit_{1}, \liabilityoneunit_{2} ) \right)^{1/2}}{\expectation_{\agerating} ( \liabilityoneunit_{1} )} \quad \textrm{as $\nummbrs \rightarrow \infty$}.
\end{displaymath}
The quantity $\left( \covariance_{\agerating} ( \liabilityoneunit_{1}, \liabilityoneunit_{2} ) \right)^{1/2} / \expectation_{\agerating} ( \liabilityoneunit_{1} )$ can be interpreted as a measure of the systematic mortality risk in the scheme.  As this quantity does not depend on the number of members $\nummbrs$, this means that under the stochastic mortality models there is systematic mortality risk in the scheme regardless of the number of members.

The remaining portion of the coefficient of variation, given by
\begin{displaymath}
\Vco_{\agerating} ( \liabilitytotal_{\nummbrs} ) -  \frac{\left( \covariance_{\agerating} ( \liabilityoneunit_{1}, \liabilityoneunit_{2} ) \right)^{1/2}}{\expectation_{\agerating} ( \liabilityoneunit_{1} )}
\end{displaymath}
is then interpreted as a measure of the idiosyncratic mortality risk in the scheme with $\nummbrs$ members.  Again, this means that when using a stochastic mortality model, the approximation $\liabilitytotal_{\nummbrs} \approx \expectation_{\agerating} ( \liabilitytotal_{\nummbrs} )$, a.s., may not be appropriate.  The appropriateness clearly depends on the magnitude of $\left( \covariance_{\agerating} ( \liabilityoneunit_{1}, \liabilityoneunit_{2} ) \right)^{1/2} / \expectation_{\agerating} ( \liabilityoneunit_{1} )$.  

The change in the coefficient of variation for various choices of $\agerating$ as the number of members increases is shown in Figure \ref{FIGVcobasic}.  For example, the blue curve in Figure \ref{FIGVcobasic} shows the results when $\agerating=1$.    For this model, when there are 100 members in the scheme, the standard deviation of the total liability value is 5.5\% of the expected value.  For a scheme with 500 members, the standard deviation of the total liability value is 4.3\% of the expected value.  In fact, even as the number of members $\nummbrs$ becomes very large, the standard deviation of the total liability value will not drop below 3.9\% of the expected value: this limit is shown by the horizontal blue line.  Thus 3.9\% of the expected value of liabilities is the amount of the standard deviation which is due to systematic risk.  This means that for the scheme with 100 members, 71\% (=$3.9 / 5.5$) of the total standard deviation is due to systematic risk and the remaining 29\% is caused by idiosyncratic risk.

The most noticeable convergence feature in Figure \ref{FIGVcobasic} is that the convergence rate of the coefficient of variation to its minimum value increases as $\agerating$ increases.  This is a consequence of the covariance term increasing at a faster rate than the variance term as the age rating $\agerating$ increases, so that the term $\variance_{\agerating} \left( \liabilityoneunit_{1} \right) -  \covariance_{\agerating} ( \liabilityoneunit_{1}, \liabilityoneunit_{2} )$ in (\ref{EQNvcoallmbr}) is decreasing in $\agerating$.  We interpret this as meaning that the idiosyncratic risk diminishes as the age rating increases.

We note that all the expected present values of the liabilities calculated using the models in Figure \ref{FIGVcobasic} are almost identical since the average age rating in each of the models is zero.  The results highlight why it may not be sufficient to value a pension scheme using different deterministic mortality bases in an analysis of the impact of uncertainty about the future distribution of deaths.  By not taking into consideration our views on the likelihood of the different mortality bases, we fail to quantify our own uncertainty about the present value of the scheme's liabilities.

As an aside, setting $\agerating=0$ in (\ref{EQNagerating}) results in the Deterministic Mortality Model.  It is clear that all the results in this subsection also hold under the Deterministic Mortality Model; in the deterministic model, lives are independent and hence $\covariance_{0} ( \liabilityoneunit_{1}, \liabilityoneunit_{2} )=0$.

\section{Effect of different benefit amounts} \label{SECexecs}

Usually in a pension scheme, members do not receive the same amount of benefits for various reasons: members have different service histories and salaries, and sections of the scheme may have different benefit structures.  Here we analyse the effect on the coefficient of variation of the members having different amounts of benefits.

\subsection{Scheme with an executive section}
Suppose that members do not all receive the same amount of benefit and a proportion $\execpropn$ of the members receive $\execmult$ units per annum at retirement, for some $\execmult \in \naturalnumbers$.  For convenience, we refer to these members as executives, but they could also represent, for example, members who are full-time employees, have a long service or have a significant transferred-in benefit.  The remaining members in the scheme receive a standard benefit of $1$ unit per annum at retirement.  Setting
\begin{displaymath}
\execfactor (\execpropn,\execmult) := \frac{\execpropn \execmult^{2} + 1-\execpropn}{(\execpropn \execmult + 1-\execpropn)^{2}},
\end{displaymath}
we find
\begin{equation} \label{EQNvcoexec}
\Vco ( \liabilitytotal_{\nummbrs} ) = \frac{1}{\expectation ( \liabilityoneunit_{1} )} \left( \frac{1}{\nummbrs} \execfactor (\execpropn, \execmult) \left( \variance \left( \liabilityoneunit_{1} \right) - \covariance ( \liabilityoneunit_{1}, \liabilityoneunit_{2} ) \right) + \covariance ( \liabilityoneunit_{1}, \liabilityoneunit_{2} ) \right)^{1/2}.
\end{equation}
We still have the result
\begin{equation} \label{EQNvcoexeclimit}
\Vco ( \liabilitytotal_{\nummbrs} ) \rightarrow  \frac{\left( \covariance ( \liabilityoneunit_{1}, \liabilityoneunit_{2} ) \right)^{1/2}}{\expectation ( \liabilityoneunit_{1} )} \quad \textrm{as $\nummbrs \rightarrow \infty$},
\end{equation}
but for finite values of $\nummbrs$ there is an extra factor $\execfactor (\execpropn,\execmult)$ in (\ref{EQNvcoexec}), which is greater than one for all $\execpropn \in [0,1]$.  The effect of $\execfactor (\execpropn,\execmult)$ is to increase the coefficient of variation compared to the scheme in which all members receive the same amount of benefit.  We examine numerically how this factor affects the coefficient of variation, particularly for smaller values of $\nummbrs$, as we vary $\execmult$.

For the numerical examples, we assume that all scheme members are age $\memberage$ and $\execpropn=5\%$ of the members receive $\execmult$ units per annum at retirement, with the remaining $1-\execpropn=95\%$ of the members receiving $1$ unit per annum at retirement.  The continuously-compounded interest rate is again 4\% per annum.

\subsubsection*{Deterministic Mortality Model}
\begin{figure}[p]
\centering
\subfigure[Deterministic Mortality Model.]
{ \label{FIGVcoexecsdet}
\includegraphics[scale=0.45]{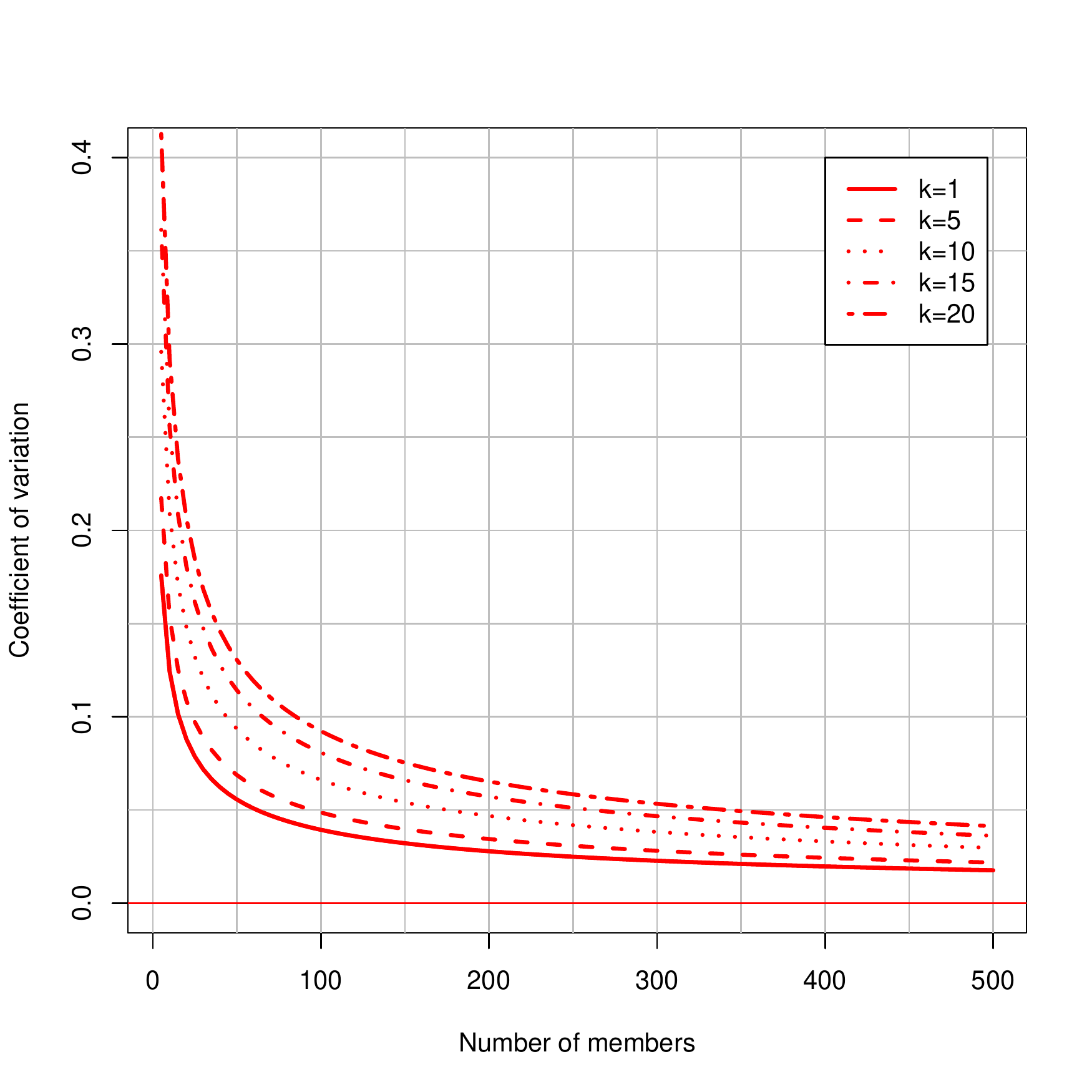} 
}
\subfigure[Stochastic Mortality Model 1]
{ \label{FIGVcoexecsstoci}
\includegraphics[scale=0.45]{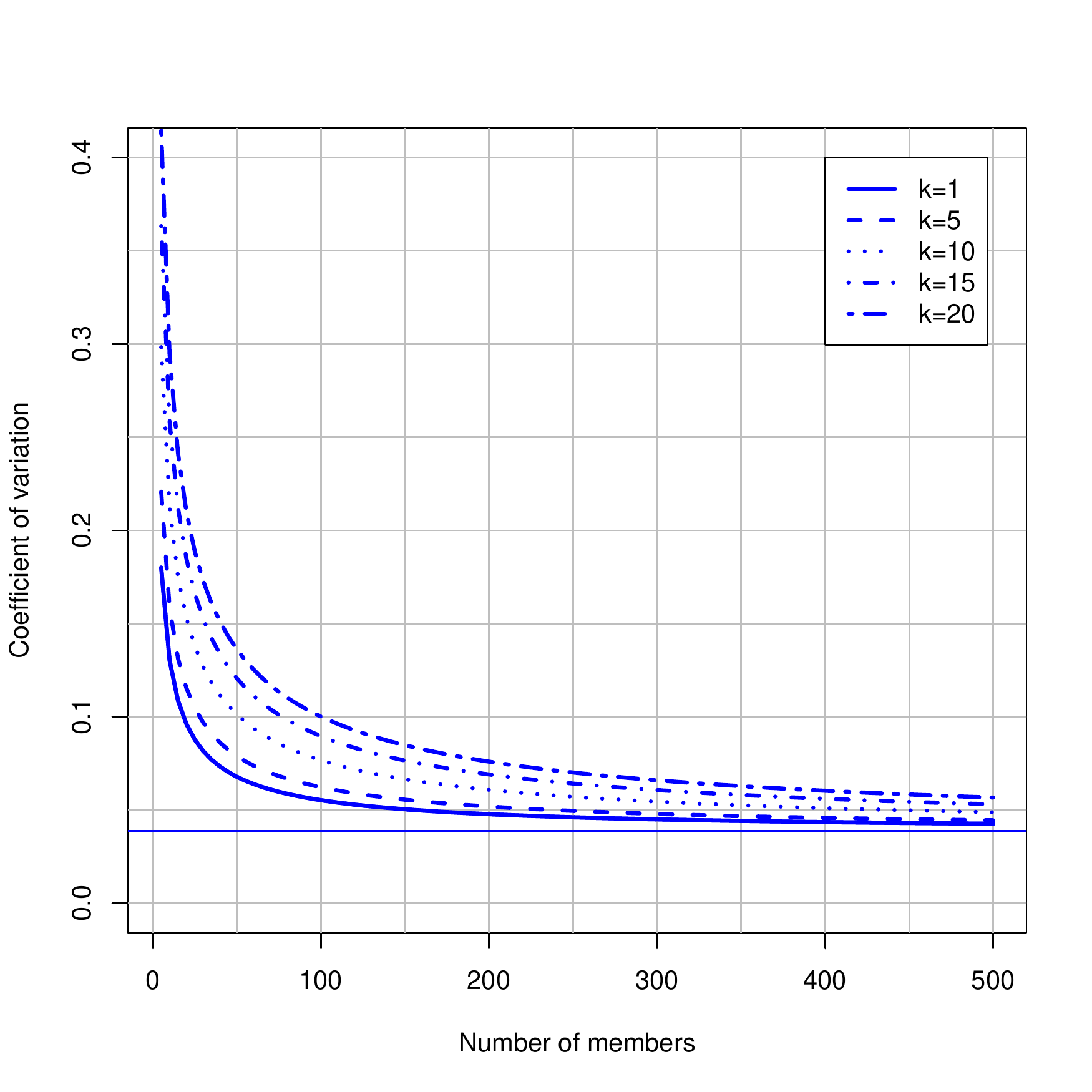} 
}
\vspace{0.2cm}
\caption{Coefficient of variation for a simple pension scheme based on the life table \lifetable and 5\% of the membership receiving $\execmult$ units per annum at retirement, with the remaining 95\% of membership receiving $1$ unit per annum.  The horizontal lines show the minimum values of the coefficient of variation that can be obtained under each mortality model.}
\label{FIGVcoexecs}
\end{figure}

First we examine the rate of convergence of the coefficient of variation under the Deterministic Mortality Model.  Under this model, lives are independent and we see from (\ref{EQNvcoexeclimit}) that the coefficient of variation converges to zero.  Figure \ref{FIGVcoexecsdet} shows how fast this convergence occurs for various values of $\execmult$.  The figure also shows a significant increase in the coefficient of variation as $\execmult$ increases.  However, all of this risk is idiosyncratic and it vanishes as the number of members increases.   Consider the scheme with 100 members.  If all members receive the same amount of benefit then the coefficient of variation is 3.9\%, which means that the standard deviation of the total liability value is 3.9\% of the expected value.  If 5 members receive $5$ times the standard benefit at retirement, then the coefficient of variation increases to 4.9\%.  With these 5 members receiving instead $20$ times the standard benefit, the coefficient of variation increases to 9.2\%.

\subsubsection*{Stochastic Mortality Model 1}
Now we consider Stochastic Mortality Model 1.  We plot in Figure \ref{FIGVcoexecsstoci} the coefficient of variation as the number of members increases for various values of $\execmult$.  Consider again the scheme with 100 members.   If all members receive the same amount of benefit then the coefficient of variation is 5.5\%.  With 5 of them receiving $5$ times the standard benefit at retirement, it increases to 6.2\%.  With the 5 executive members receiving instead $20$ times the standard benefit, it becomes 10.0\%.

Some of this risk is systematic: the horizontal blue line in Figure \ref{FIGVcoexecsstoci} shows the minimum coefficient of variation for this model and is a measure of the amount of systematic risk in each scheme.  The amount of coefficient of variation above this line is a measure of the idiosyncratic risk in each scheme.  We see clearly in the figure the reduction in idiosyncratic risk as the total number of scheme members $\nummbrs$ increases.   

\subsection{Summary}
Having a relatively small proportion of the membership receiving elevated benefits can increase the idiosyncratic mortality risk of the scheme considerably, particularly for smaller schemes.  Even if we are confident about the future distribution of deaths of the scheme members, the mortality risk of the scheme, as measured by the coefficient of variation, is significantly higher when there is a small executive section.  The simple reason for this is that paying 10 units of benefit to one member is more risky that paying 1 unit of benefit to 10 members.

The natural question to ask as a result of these observations is: can we quantify the risk posed by the executive section to the overall risk of the scheme?  The answer to this question can help us to decide if it is reasonable to buy-out or insure the executive section's benefits in order to manage the scheme's mortality risk.  To this end, we turn to a capital allocation idea to attempt to analyse the risk concentrations in the executive section.

\section{Risk capital allocation} \label{SECcapitalallocation}

Here we use the standard deviation of the total liability value $\stddev \left( \liabilitytotal_{\nummbrs} \right)$ as a measurement of the mortality risk in the pension scheme.  We seek to allocate $\stddev \left( \liabilitytotal_{\nummbrs} \right)$ to each member, or group of members, to reflect their individual contribution to the risk in the pension scheme.  We call $\stddev \left( \liabilitytotal_{\nummbrs} \right)$ the \emph{risk capital}.  It is important to realise that this allocation must be considered in the context of the entire scheme so that we can allow for the impact of the dependencies between scheme members' future lifetimes on the overall mortality risk of the scheme.

We use the \emph{Euler capital allocation principle} to allocate the risk capital between members.  The theoretical details of the Euler capital allocation principle can be found in \cite[Section 6.3.2]{mcneiletal.book}.  We set out next how this principle is applied to the pension scheme.

The idea is that we wish to find amounts $\riskweight_{1}, \ldots, \riskweight_{\nummbrs}$, corresponding to each of the scheme members, such that
\begin{displaymath}
\sum_{n=1}^{\nummbrs} \riskweight_{n} = \stddev \left( \liabilitytotal_{\nummbrs} \right).
\end{displaymath}
We call $\riskweight_{n}$ the risk capital for member $n$ and we interpret it as the amount of the total standard deviation in the pension scheme which can be attributed to member $n$.    Recall that the random variable $\liability_{n}$ denotes the present value of the benefit payable to member $n$ and the present value of the total liability is
\begin{displaymath}
\liabilitytotal_{\nummbrs} = \sum_{n=1}^{\nummbrs} \liability_{n}.
\end{displaymath}
The Euler capital allocation principle tells us that, when the standard deviation is used as the risk measure, we can calculate
\begin{displaymath}
\riskweight_{n} = \frac{\covariance ( \liability_{n}, \liabilitytotal_{\nummbrs})}{\stddev \left( \liabilitytotal_{\nummbrs} \right)} = \frac{ \sum_{m=1}^{\nummbrs} \covariance ( \liability_{n},\liability_{m})}{\stddev \left( \liabilitytotal_{\nummbrs} \right)}.
\end{displaymath}
From the latter formula it is clear why the Euler capital allocation principle is also called the \emph{covariance principle}.

\subsection{Homogeneous pension scheme}

We take the example where all members receive the same retirement benefit $\benefit_{n} = 1$, for all $n$.  Then as all members are assumed to be the same age,
\begin{displaymath}
\riskweight_{n} = \frac{1}{\nummbrs} \stddev \left( \liabilitytotal_{\nummbrs} \right) = \left( \frac{1}{\nummbrs} \left( \variance \left( \liabilityoneunit_{1} \right) - \covariance ( \liabilityoneunit_{1}, \liabilityoneunit_{2} ) \right) +  \covariance ( \liabilityoneunit_{1}, \liabilityoneunit_{2} )\right)^{1/2}.
\end{displaymath}
As expected, in a scheme where all members receive exactly the same benefit and are the same age, then all members are allocated the same amount of risk capital.  Furthermore, 
\begin{displaymath}
\riskweight_{n} \rightarrow \left( \covariance ( \liabilityoneunit_{1}, \liabilityoneunit_{2} ) \right)^{1/2} \quad \textrm{as $\nummbrs \rightarrow \infty$}. 
\end{displaymath}
We interpret $\left( \covariance ( \liabilityoneunit_{1}, \liabilityoneunit_{2} ) \right)^{1/2}$ as the amount of systematic mortality risk and
$\riskweight_{n} - \left( \covariance ( \liabilityoneunit_{1}, \liabilityoneunit_{2} ) \right)^{1/2}$ as the amount of idiosyncratic mortality risk for member $n$.
For example, under the Deterministic Mortality Model, $\riskweight_{n} \rightarrow 0$ as  $\nummbrs \rightarrow \infty$ and all of the risk capital allocation $\riskweight_{n}$ can be interpreted as the amount of idiosyncratic mortality risk for member $n$.

\subsection{Pension scheme with an executive section}
Now suppose that there is an executive section in the scheme, as in Section \ref{SECexecs}, so that a proportion $\execpropn$ of the members receive $\execmult$ units per annum at retirement, for some $\execmult \in \naturalnumbers$.  All other members receive the standard benefit of $1$ unit per annum at retirement.

We seek to allocate the risk capital between the executive and non-executive section and analyse the split between systematic and idiosyncratic risk.  First, we examine the risk capital allocated to each member type, executive and non-executive.  Denoting the risk capital allocated to an executive member by $\riskweightexec$ and the amount allocated to a non-executive member by $\riskweightnormal$, we find that 
\begin{displaymath}
\riskweightexec \rightarrow \execmult \left( \covariance ( \liabilityoneunit_{1}, \liabilityoneunit_{2} ) \right)^{1/2} \quad \textrm{and} \quad \riskweightnormal \rightarrow \left( \covariance ( \liabilityoneunit_{1}, \liabilityoneunit_{2} ) \right)^{1/2} \quad \textrm{as $\nummbrs \rightarrow \infty$}. 
\end{displaymath}
Interpreting the limits as the amount of systematic risk per member type, Table \ref{TABcapitalalloc} shows the risk capital allocated by member type and split by mortality risk component.  Summing the limits over the scheme membership, then we identify
\begin{equation} \label{EQNtotaysysmort}
\nummbrs (\execpropn \execmult + 1 - \execpropn) \left( \covariance ( \liabilityoneunit_{1}, \liabilityoneunit_{2} ) \right)^{1/2}
\end{equation}
as the total systematic mortality risk in the scheme.   The proportion of the systematic mortality risk which is allocated to the executive section is
\begin{displaymath}
\riskallocbenefitexec := \frac{\execpropn \nummbrs \execmult \left( \covariance ( \liabilityoneunit_{1}, \liabilityoneunit_{2} ) \right)^{1/2}}{\nummbrs (\execpropn \execmult + 1 - \execpropn) \left( \covariance ( \liabilityoneunit_{1}, \liabilityoneunit_{2} ) \right)^{1/2}} = \frac{\execpropn \execmult}{\execpropn \execmult + 1 - \execpropn}.
\end{displaymath}
The latter interpretation of $\riskallocbenefitexec$ only makes sense when there is non-zero systematic mortality risk.  For example, under the Deterministic Mortality Model, $\covariance ( \liabilityoneunit_{1}, \liabilityoneunit_{2} )=0$ and there is no systematic mortality risk.  Note that when there is systematic mortality risk, the proportion $\riskallocbenefitexec$ allocated to the executive section is invariant to the choice of the mortality model (although the risk capital per executive member depends on the mortality model; see Table \ref{TABcapitalalloc}).
{
\renewcommand{\arraystretch}{1.3}
 \begin{table}
 \caption{Allocation using the Euler capital allocation principle of the risk capital by member and mortality risk type.}
  \centering
  \begin{tabular}{ | l | c | c | }
    \hline
   & \multicolumn{2}{| c |}{Mortality risk component} \\ \hline
  Member type & Systematic & Idiosyncratic \\ \hline
 Non-executive & $\left( \covariance ( \liabilityoneunit_{1}, \liabilityoneunit_{2} ) \right)^{1/2}$ & $\riskweightnormal - \left( \covariance ( \liabilityoneunit_{1}, \liabilityoneunit_{2} ) \right)^{1/2}$ \\
  Executive & $\execmult \left( \covariance ( \liabilityoneunit_{1}, \liabilityoneunit_{2} ) \right)^{1/2}$ & $\riskweightexec - \execmult \left( \covariance ( \liabilityoneunit_{1}, \liabilityoneunit_{2} ) \right)^{1/2}$ \\
    \hline
  \end{tabular}
\label{TABcapitalalloc}
\end{table}
}

\begin{figure}[p]
\centering
\subfigure[$\nummbrs=100$.]
{ \label{FIGeulerk5N100}
\includegraphics[scale=0.42]{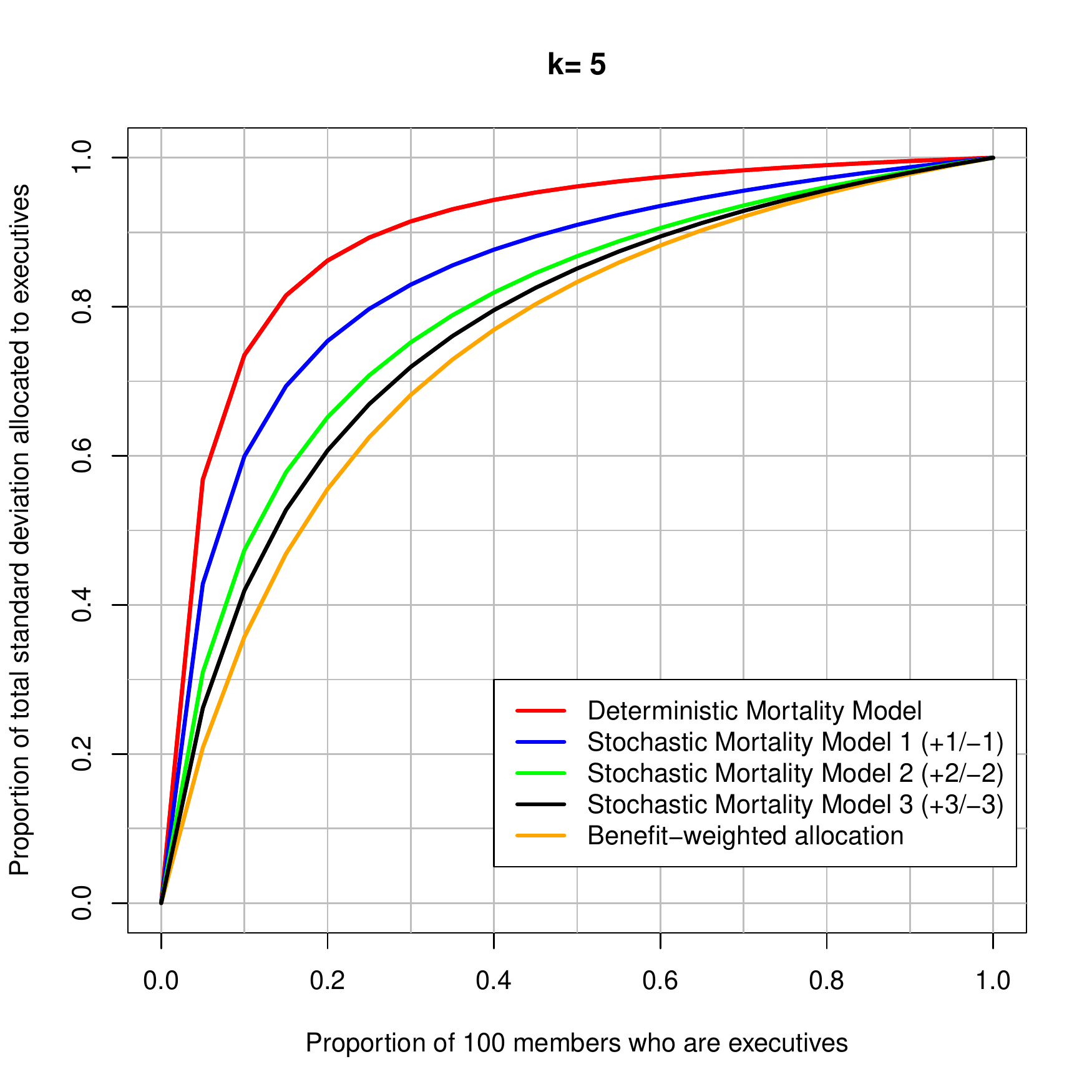} 
}
\subfigure[$\nummbrs=500$]
{ \label{FIGeulerk5N500}
\includegraphics[scale=0.42]{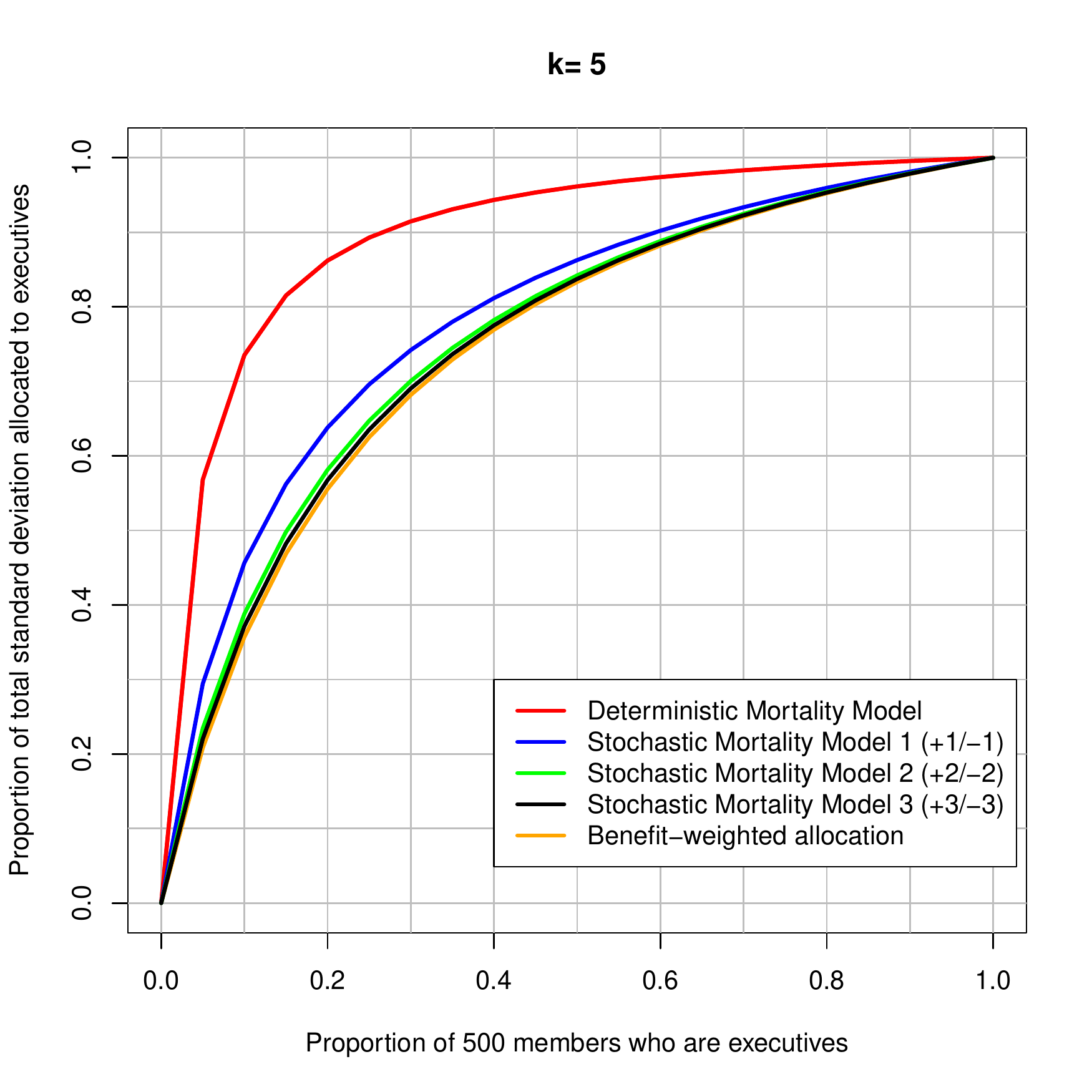} 
}
\vspace{0.2cm}
\caption{Proportion of the total standard deviation allocated using the Euler capital allocation principle to the executive section of a pension scheme with an executive section receiving 5 times the benefit of the non-executive section.  In Figure \ref{FIGeulerk5N100}, the total number of scheme members is $\nummbrs=100$ and in Figure \ref{FIGeulerk5N500}, $\nummbrs=500$.  The red and orange curves are identical in both figures. }
\label{FIGeulerk5}
\end{figure}

\begin{figure}[p]
\centering
\subfigure[$\nummbrs=100$.]
{ \label{FIGeuleralpha5N100}
\includegraphics[scale=0.45]{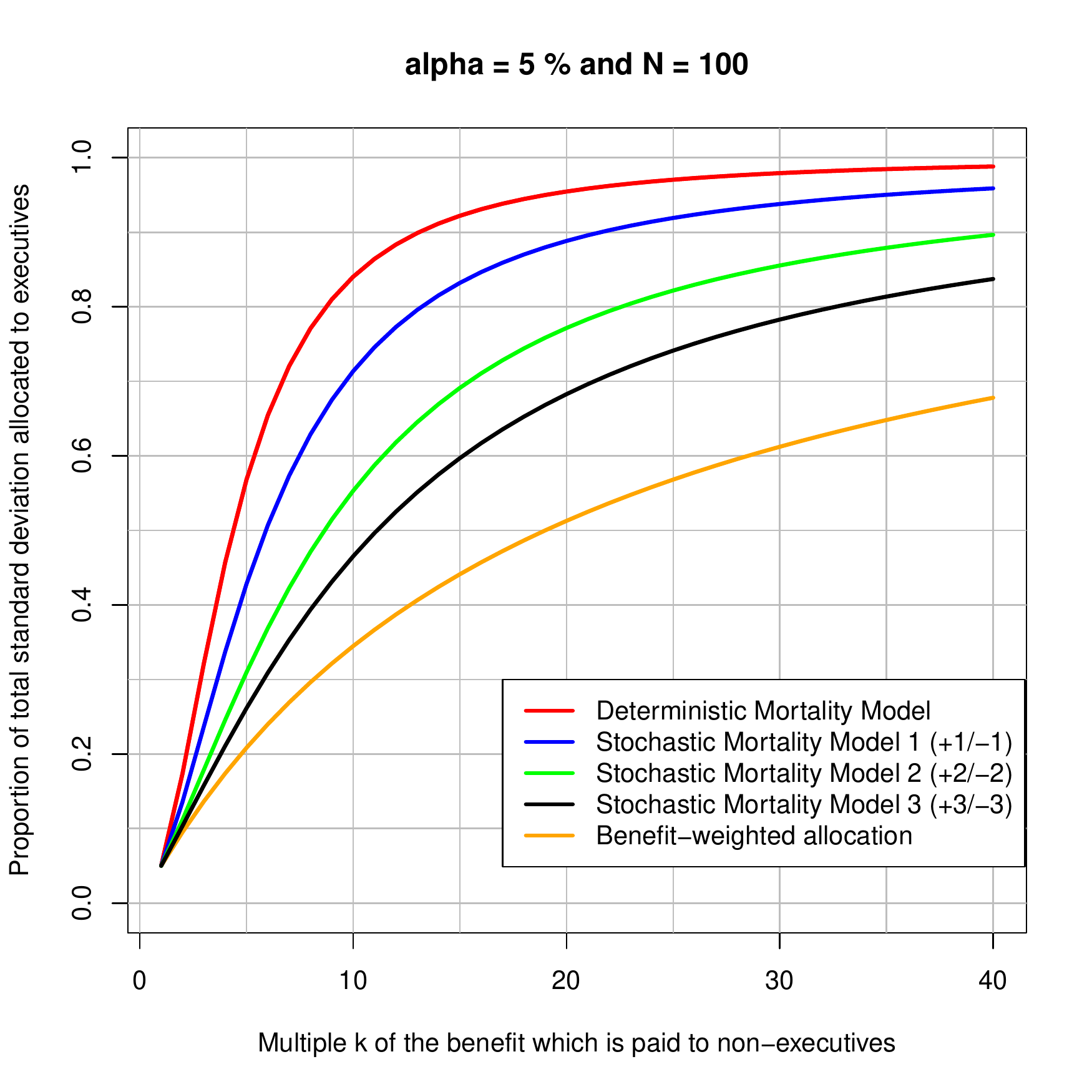} 
}
\subfigure[$\nummbrs=500$.]
{ \label{FIGeuleralpha5N500}
\includegraphics[scale=0.45]{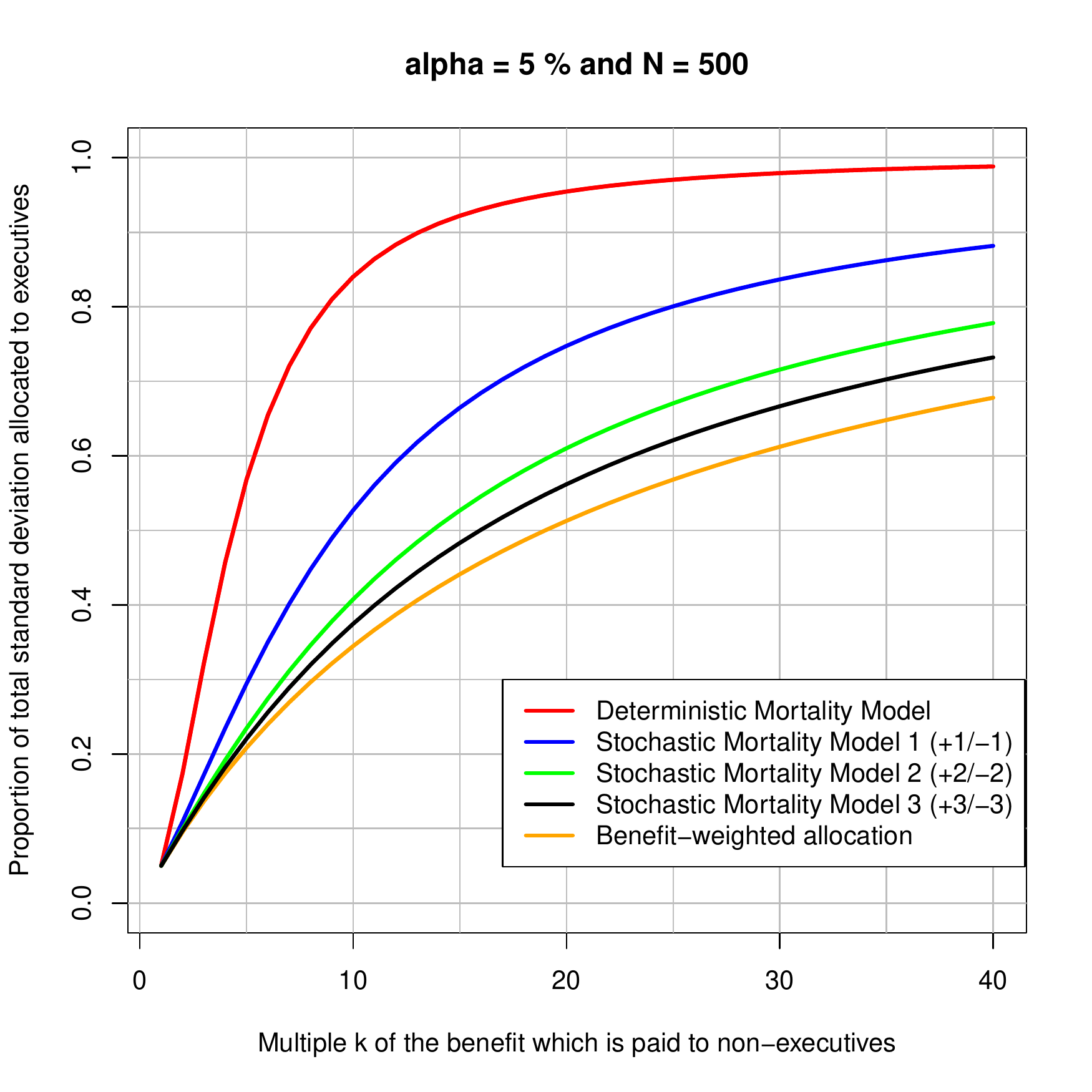} 
}
\vspace{0.2cm}
\caption{Proportion of the total standard deviation allocated using the Euler capital allocation principle to the executive section of a pension scheme with an executive section constituting 5\% of the total membership.  In Figure \ref{FIGeuleralpha5N100}, the total number of scheme members is $\nummbrs=100$ and in Figure \ref{FIGeuleralpha5N500}, $\nummbrs=500$.  The red and orange curves are identical in both figures. }
\label{FIGeuleralpha5}
\end{figure}

We can also interpret $\riskallocbenefitexec$ as the proportion of the risk capital allocated to the executive section using a \emph{benefit-weighted approach}.  The benefit-weighted approach allocates the risk capital between members using their benefit amounts as weights.  The orange curves in Figures \ref{FIGeulerk5} and \ref{FIGeuleralpha5} show how $\riskallocbenefitexec$ varies as $\execpropn$ and $\execmult$ vary, respectively.

While the executive members contribute $\execmult$ times the systematic mortality risk of the non-executives, they contribute more to the idiosyncratic mortality risk.  This can be seen by examining the relation
\begin{equation} \label{EQNriskweightexec}
\riskweightexec = \execmult \riskweightnormal + \execmult \left( \execmult - 1 \right) \frac{\left( \variance(\liabilityoneunit_{1}) - \covariance(\liabilityoneunit_{1}, \liabilityoneunit_{2}) \right)}{\stddev \left( \liabilitytotal_{\nummbrs} \right)}.
\end{equation}
For $\execmult=1$, the expression collapses to $\riskweightexec = \riskweightnormal$.  For $\execmult>1$, the risk capital allocated to the executive members is greater than $\execmult$ times the risk capital allocated to the non-executive members, even though the benefit paid to the executive members is $\execmult$ times the benefit paid to the non-executives.  However, as $\nummbrs$ increases, the risk capital $\stddev ( \liabilitytotal_{\nummbrs} )$ also increases and the magnitude of the second term on the right-hand side of (\ref{EQNriskweightexec}) tends to zero.  Hence $\riskweightexec \rightarrow \execmult \riskweightnormal$ as $\nummbrs \rightarrow \infty$ and, in the limit, all the mortality risk is systematic risk as the idiosyncratic risk vanishes.  Thus the second term on the right-hand side of (\ref{EQNriskweightexec}) is the amount of idiosyncratic risk allocated to each executive over and above $\execmult$ times the amount of idiosyncratic risk allocated to each non-executive.

Next we allocate the risk capital between the executive and non-executive sections.  As
\begin{displaymath}
\stddev \left( \liabilitytotal_{\nummbrs} \right) = \execpropn \nummbrs \cdot \riskweightexec + (1 - \execpropn) \nummbrs \cdot \riskweightnormal,
\end{displaymath}
then
\begin{displaymath}
\riskallocexec := \execpropn \nummbrs \frac{ \riskweightexec}{\stddev \left( \liabilitytotal_{\nummbrs} \right)}
\end{displaymath}
is the proportion of the risk capital allocated to the executive section using the Euler capital allocation principle.  Figures \ref{FIGeulerk5} and \ref{FIGeuleralpha5} show how $\riskallocexec$ varies as $\execpropn$ and $\execmult$ vary, respectively, under various mortality models.  For the figures, we assume that all scheme members are age $\memberage$ and the continuously-compounded interest rate is again 4\% per annum.  Note that for the special case where we use the Deterministic Mortality Model, we find
\begin{equation} \label{EQNdmmriskalloc}
\riskallocexec = \frac{\execpropn \execmult^2}{\execpropn \execmult^2 + 1 - \execpropn},
\end{equation}
so that for our simple defined-benefit pension scheme, where all members are the same age, the Euler capital allocation to the executive section is invariant with respect to the total number of members.  Hence the red curves in Figure \ref{FIGeulerk5} are identical, and the red curves in Figure \ref{FIGeuleralpha5} are also identical.  However, this is not true in general.

The two figures show that the proportion of the risk capital allocated to the executive section is higher using the Euler capital allocation principle than using a benefit-weighted approach.  For the Stochastic Mortality Models, as the number of members increases in a scheme, the idiosyncratic risk decreases and the capital allocation using the Euler capital allocation principle converges to the benefit-weighted approach allocation; this result can be seen in Figures \ref{FIGeulerk5} and \ref{FIGeuleralpha5}.  Note that the convergence rate is greatest in the models where the difference in the age rating $\agerating$ is greatest, for the same reasons as given in Subsection \ref{SUBSECsmm} for the convergence observed in Figure \ref{FIGVcobasic}.

If we instead interpret the orange curves in Figures \ref{FIGeulerk5} and \ref{FIGeuleralpha5} as the proportion of the systematic mortality risk allocated to the executive section, then we deduce that the executive section contributes a greater proportion to the total idiosyncratic mortality risk than to the total systematic mortality risk of the scheme (although the amount of the idiosyncratic risk may be much smaller than the amount of systematic risk), consistent with our earlier observation.

\subsection{Summary}
Risk capital allocation is a way of identifying sections of the scheme where risk is concentrated.  Using the Euler capital allocation principle to allocate the standard deviation of the total liability value between sections, there appears to be higher concentration of idiosyncratic mortality risk in the executive section than would be suggested by a benefit-weighted allocation.  Thus, particularly for small pension schemes, the mortality risks in the executive section should be carefully measured and managed.

\section{Summary} \label{SECsummary}
The idiosyncratic mortality risk in pension schemes is often ignored by pensions actuaries since it is assumed to be inconsequential.  However, as we demonstrate, it can be significant even for schemes with a few hundred members.  While the effect of the idiosyncratic mortality risk reduces in the stochastic mortality models we examine, this is at a cost of increasing the overall mortality risk.  The conclusion is that, to manage appropriately the risks facing a pension scheme, it is important to quantify the mortality risk since it should not be assumed to be negligible.

We also investigate allocating the standard deviation of the present value of the total liabilities of the scheme between an executive and non-executive section of the scheme.  Since the financial assumptions of the scheme are deterministic, we regard the standard deviation as a measure of the mortality risk in the scheme.  Thus the allocation is a way of seeing where mortality risk may be concentrated in the scheme.  Using the Euler capital allocation principle, we find that there is a concentration of mortality risk in the executive section.  In particular, the amount of the total idiosyncratic mortality risk allocated to the executives is much higher than would be suggested by their benefit amount.  Thus it is not enough to measure the overall mortality risk of the scheme, but also to do this within the scheme.  This allows the pension scheme to manage the mortality risks of the scheme, for example by aiding in the decision to purchase annuities in from an insurance company in respect of all or part of the liability in order to reduce the scheme's mortality risk.

Furthermore, while we have done the analysis in the context of small defined-benefit pension schemes, much of it is relevant to large pension schemes and to insurance companies selling life annuities of different benefit amounts.

Having analysed the mortality risk in isolation, the effect of other risks, such as financial risks, on the standard deviation and coefficient of variation of the liability remains to be studied.  The impact of risk mitigation investment strategies on the financial health of the scheme can also be studied, which, very importantly, introduces the scheme's assets into the analysis.

\bibliographystyle{plainnat}
\bibliography{article}

\end{document}